\documentstyle[prl,aps,epsf,floats]{revtex}
\begin{document}
\twocolumn[
\hsize\textwidth\columnwidth\hsize\csname@twocolumnfalse\endcsname
\draft
]
\noindent{\bf 
Comment on "Fluctuation-Driven First-Order Transition in Pauli-Limited $d$-Wave Superconductors":}

\bigskip

\ Recently, the presence of a kind of Fulde-Ferrell-Larkin-Ovchinnikov (FFLO) state in a heavy fermion superconductor CeCoIn$_5$ at low temperatures in a field ${\bf H}$ parallel to the superconducting layers (${\bf H} \perp c$) has been clarified through thermodynamic \cite{LANL,FSU} and ultrasound \cite{ISSP} measurements. Dalidovich and Yang (DY) \cite{DY} have argued by assuming the Pauli-limiting, i.e., neglecting the presence of vortices, that a fluctuation-driven first order transition (FOT) resulting from their perturbative renormalization-group analysis be consistent with the apparent FOT near $H_{c2}(T)$ between the normal and FFLO states in CeCoIn$_5$. 
On the other hand, the observed anomaly in the sound velocity seems to be consistent only with a change of tilt elasticity of vortices due to a growth of a modulation parallel to ${\bf H}$ \cite{ISSP}. 

Below, we point out that the DY's neglect of vortices is never justified in addressing the phase diagram of CeCoIn$_5$. First, the anisotropy $\gamma$ in coherence lengths of CeCoIn$_5$ is so small that the orbital depairing inducing the vortices is not negligible even in the mean field (MF) analysis. The $\gamma$-value estimated from $H_{c2}$-data \cite{SI} is at most 2.4, and hence, CeCoIn$_5$ is less anisotropic than YBa$_2$Cu$_3$O$_{6-\delta}$ with $\gamma \geq 5$ and $\xi_0/\gamma < s$, where $\xi_0$ and $s$ are the in-plane coherence length and the spacing between neighboring layers, respectively. Both materials have quasi two-dimensional electronic states. However, the phase diagram of YBa$_2$Cu$_3$O$_{6-\delta}$ in ${\bf H} \perp c$ is explained as that of Josephson vortex (JV) states \cite{RIpara}, implying the importance of interlayer couplings, while even an entire confinement of vortices in the interlayer spacings cannot occur in CeCoIn$_5$ with $\xi_0/\gamma > s$. 

Next, DY assume the normal to FFLO transition at $H_{c2}(T)$ to be of second order in the MF approximation. The FOT-like behaviors near $H_{c2}$ in CeCoIn$_5$ \cite{Izawa,LANL} are already visible far above the temperature $T_{\rm FFLO}$ below which $H_{c2}(T)$ is the MF boundary between the normal and FFLO states. In addition, a slight hysteresis \cite{LANL} has appeared just {\it above} $T_{\rm FFLO}$ in sweeping the temperature and hence, has nothing to do with the presence of the FFLO state below $T_{\rm FFLO}$. Under the circumstances, assuming \cite{DY} a second order MF transition at $H_{c2}(T)$ {\it only} below $T_{\rm FFLO}$ is clearly unreasonable. 

Once including the fluctuation, the neglect of the vortices becomes more serious. The melting of JV solid is a weak FOT, which easily becomes continuous due to a pinning disorder, and there is no phase coherence in the resulting JV liquid \cite{RIpara} below the (sharp) crossover line $H_{c2}(T)$. This picture on the vortex phase diagram is unaffected by the character of the MF transition at $H_{c2}$ and a FFLO modulation which, due to the orbital depairing, tends to be formed along ${\bf H}$ \cite{AI}. In the Pauli-limited case, however, the noncritical FOT \cite{DY} should occur just {\it above} $H_{c2}$ because it follows by assuming $H_{c2}$ as a possible critical field, and the FFLO modulation direction is not fixed by ${\bf H}$. That is, since the fluctuation effect in the case with the orbital depairing is incompatible with that in the Pauli-limited case, assuming the Pauli-limiting is not valid even qualitatively in CeCoIn$_5$ showing the vortex elasticity below $T_{\rm FFLO}$ \cite{ISSP}. 
A slight hysteresis seen in ${\bf H} \perp c$ only at low temperatures and the FOT-like behavior in CeCoIn$_5$ have been explained within the approach including both the orbital and spin depairings \cite{AI}. 

Further, according to the thermal conductivity data \cite{Tanatar,Capan}, the normal to FFLO crossover at $H_{c2}$ in an organic superconductor $\lambda$-(BETS)$_2$GaCl$_4$, where a convincing sign of a FFLO transition below $H_{c2}$ was found \cite{Tanatar}, is clearly continuous. If the fluctuation of this material be much weaker than that of CeCoIn$_5$, this might not contradict the DY's FOT based on a continuous MF transition. However, judging from an experimental phase diagram \cite{Mielke} in ${\bf H} \parallel c$ which is similar to those of $\kappa$-(ET)$_2$-salts examined elsewhere \cite{RI}, $\lambda$-(BETS)$_2$GaCl$_4$ has a much stronger fluctuation than CeCoIn$_5$ in spite of its $\gamma$-value ($\simeq 5$) \cite{Tanatar} in magnitude close to that of CeCoIn$_5$. As indicated elsewhere \cite{AI}, such a continuous normal to FFLO crossover \cite{Tanatar} is naturally explained as a fluctuation effect once taking account of the orbital depairing. 


\vspace{0.3truecm}
\bigskip

\noindent
Ryusuke Ikeda$^1$ and Hiroto Adachi$^2$

\noindent
$^1$Department of Physics, 
Kyoto University, Kyoto 606-8502, Japan \\
$^2$Department of Physics, Okayama University, Okayama 700-8530, 
Japan

\vspace{0.3mm}

\end{document}